\documentclass[aps, prx, twocolumn, floatfix, superscriptaddress]{revtex4-2}

\usepackage{amsmath,amssymb,amsfonts,amsbsy}
\usepackage{graphicx}
\usepackage{subfig}
\usepackage{dcolumn}
\usepackage{bm}
\usepackage{multirow}
\usepackage{mathtools}
\usepackage{array}
\usepackage{color}
\usepackage[normalem]{ulem}
\usepackage[per-mode=symbol]{siunitx}
\usepackage{upgreek}
\usepackage{esvect}
\usepackage{booktabs}
\usepackage[utf8]{inputenc}
\usepackage[T1]{fontenc}
\usepackage{lmodern}
\usepackage[pagewise]{lineno}
\usepackage[labelfont=bf]{caption}
\usepackage[figurename=Fig.]{caption}
\usepackage{etoolbox}
\DeclareSIUnit \belm {Bm}
\usepackage{todonotes}
\usepackage{braket}
\usepackage{caption}
\usepackage{float}
\usepackage{tabularx}

\usepackage[hidelinks,bookmarks=false]{hyperref} 
\usepackage[capitalise,nameinlink]{cleveref} 

\newcommand{\xiao}[1]{\textcolor{blue}{#1}}

\def\@setaltaffiliation{\vspace{-\baselineskip}\def\altaffiliation##1{\@par##1\@addpunct.}\altaffiliationes}
\def\@setaltaffiliation{\vspace{-\baselineskip}\def\altaffiliation##1{\@par##1\@addpunct.}\altaffiliationes}

\let\oldequation\align
\let\oldendequation\endalign

\renewenvironment{align}
  {\linenomathNonumbers\oldequation}
  {\oldendequation\endlinenomath}

\def\@setaltaffiliation{\vspace{-\baselineskip}\def\altaffiliation##1{\@par##1\@addpunct.}\altaffiliationes}
\def\@setaltaffiliation{\vspace{-\baselineskip}\def\altaffiliation##1{\@par##1\@addpunct.}\altaffiliationes}

\begin{document}

\newcolumntype{L}[1]{>{\raggedright\arraybackslash}p{#1}}
\newcolumntype{C}[1]{>{\centering\arraybackslash}p{#1}}
\newcolumntype{R}[1]{>{\raggedleft\arraybackslash}p{#1}}

\title{Controllable interaction between photons and distant spins \\via vacuum Rabi oscillations}
\author{Xiao~Xue}
\thanks{These authors contributed equally to this work}
\altaffiliation{\\Correspondence emails: \\xiao.xue@ustc.edu.cn; l.m.k.vandersypen@tudelft.nl}
\author{Jurgen~Dijkema}
\thanks{These authors contributed equally to this work}
\altaffiliation{\\Correspondence emails: \\xiao.xue@ustc.edu.cn; l.m.k.vandersypen@tudelft.nl}
    \affiliation{QuTech and Kavli Institute of Nanoscience, Delft University of Technology, Lorentzweg 1, 2628 CJ Delft, Netherlands}

\author{Tobias~Bonsen}
\author{Patrick~Harvey-Collard}
\affiliation{QuTech and Kavli Institute of Nanoscience, Delft University of Technology, Lorentzweg 1, 2628 CJ Delft, Netherlands}

\author{Maximilian~Rimbach-Russ}
\affiliation{QuTech and Kavli Institute of Nanoscience, Delft University of Technology, Lorentzweg 1, 2628 CJ Delft, Netherlands}

\author{Sander~L.~de~Snoo}
\affiliation{QuTech and Kavli Institute of Nanoscience, Delft University of Technology, Lorentzweg 1, 2628 CJ Delft, Netherlands}

\author{Guoji~Zheng}
\affiliation{QuTech and Kavli Institute of Nanoscience, Delft University of Technology, Lorentzweg 1, 2628 CJ Delft, Netherlands}

\author{Amir~Sammak}
\affiliation{QuTech and Netherlands Organization for Applied Scientific Research (TNO), Stieltjesweg 1, 2628 CK Delft, Netherlands}

\author{Giordano~Scappucci}
\author{Lieven~M.~K.~Vandersypen}
\affiliation{QuTech and Kavli Institute of Nanoscience, Delft University of Technology, Lorentzweg 1, 2628 CJ Delft, Netherlands}


\date{\today}

\begin{abstract}

Vacuum Rabi oscillations between a single photon and a single spin demonstrate the capability of harnessing light-matter interaction at the level of a single quantum of energy. Since the observation of strong spin-photon coupling in gate-defined quantum dots~\cite{@samkharadze2018, @mi2018, @landig2018}, probing this interaction in the time-domain has been a major objective. Here, we carefully engineer a device composed of two spatially separated double quantum dots hosting single electron spin qubits and a superconducting cavity to accommodate microwave photons~\cite{@collard2022}. We observe multiple vacuum Rabi oscillations between each spin qubit and the cavity. By concatenating vacuum Rabi oscillations involving the two spins, an energy excitation in one qubit can be emitted as a photon and then transferred to the other qubit~\cite{sillanpaa2007coherent}. When a single photon is emitted, the cavity is prepared in a Fock state, leading to an accelerated vacuum Rabi frequency~\cite{hofheinz2008generation}. These results serve as building blocks not only in exploring light-matter interactions, but also in interfacing semiconductor spin qubits to photonic links.
\end{abstract}

\maketitle

\section{Introduction}

\begin{figure*}[htbp] 
\center{\includegraphics[width=\linewidth]{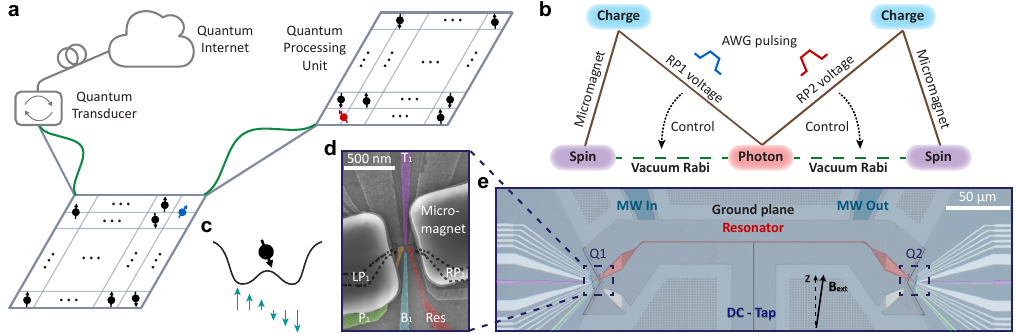}}
\caption{A distributed quantum computation architecture enabled by superconducting resonators. \textbf{a.} Vision of a spin-based quantum processor. The processor contains multiple two-dimensional quantum dot arrays with densely patterned quantum dots that host electron (or hole) spin qubits. Each module is connected to superconducting resonators at its corners. The quantum information encoded in spin qubits can be exchanged into microwave photons in the resonators, which can either be transferred to another qubit in a different module, or transduced to an optical photon in a quantum internet. \textbf{b.} Spin-photon-spin coupling scheme. Green and brown lines represent resonant and off-resonant coupling respectively. Direct couplings, i.e. the charge-photon electric-dipole coupling and the spin-charge coupling enabled by the micromagnet gradients, are indicated by solid lines. Indirect couplings, i.e. the spin-photon couplings are indicated by dashed lines. \textbf{c.} Schematic of a double dot flopping-mode spin qubit sitting in a transverse magnetic field gradient. \textbf{d.} Scanning electron microscope (SEM) image of a DQD nominally identical to the one used to host Q1. The gates P$_1$ and Res accumulate an electron in the DQD, with Res connected to the resonator and where T$_1$ and B$_1$ control the interdot tunnel coupling. RP$_1$ and LP$_1$, mostly hidden under the micromagnet, are pulsed to control the detuning and induce EDSR respectively. This panel is rotated by 90 degrees clockwise with respect to the device image (\textbf{e}). \textbf{e.} False-colored optical image of a device like the one used, which shows the resonator and the gate fan-out of the DQDs. The Res gate, which is galvanically connected to the resonator, is biased via the DC-tap. Microwave signals are sent through the in/out ports to dispersively probe the states of the qubits. The arrow is indicative of the orientation of the external magnetic field.}
\label{fig:device}
\end{figure*}

Light-matter interaction taken to the extreme quantum limit involves the coherent coupling of a single matter qubit and a single photon. When these two systems are resonant with each other, a coherent exchange of excitations between them can be observed, provided the mutual coupling strength exceeds their respective decoherence rates. In practice, this requires photons confined in a resonant cavity with a high quality factor, and a matter qubit that exhibits a large electric dipole moment and long-lived quantum coherence. In this case, when the system evolves in time starting from an empty cavity and the matter qubit in the excited state, so-called vacuum Rabi oscillations result. In addition to their fundamental interest, vacuum Rabi oscillations could also enable quantum information processing protocols and scaling of quantum computers, exploiting real photons as ``flying qubits''~\cite{divincenzo2000physicalim, georgescu2020divincenzo} to interconnect separate computational modules~\cite{@monroe2016}. These modules can reside on the same chip~\cite{sillanpaa2007coherent} or, when microwave photons are converted into  optical photons via a quantum transducer~\cite{mirhosseini2020superconducting, arnold2023all}, on separate chips and even across different qubit platforms~(\cref{fig:device}a).

Vacuum Rabi oscillations have been observed for a variety of atomic and superconducting qubits using either optical or microwave cavities~\cite{Brune1996quantum, Johansson2006vacuum, bose2014all}. To date, however, vacuum Rabi oscillations involving a single spin have remained out of reach, despite a sustained effort in this direction for many years. A promising route to this goal entails electron or hole spins confined in electrostatically defined semiconductor double quantum dots, capacitively coupled to an on-chip microwave superconducting resonator. The strong-coupling regime between a single spin qubit and a single microwave photon, a prerequisite for vacuum Rabi oscillations, was reached for the first time almost ten years ago~\cite{@mi2018,@samkharadze2018,@landig2018}. Since then, the spin-photon coupling strength has been considerably enhanced~\cite{@yu2023}, and spins in two distant double dots have been coupled strongly to the same resonator~\cite{@borjans2019,@collard2022}. The dispersive regime has also been explored, whereby the spin and photon are detuned in energy by much more than their coupling strength~\cite{@benito2019}. Dispersive readout of a single spin was demonstrated~\cite{@mi2018, dijkema2025cavity}, albeit requiring extensive averaging, and spectroscopic measurements revealed the interaction between two distant spins, dispersively coupled by virtual photons in the resonator~\cite{@collard2022}. Most recently, this interaction could be harnessed in the time domain, producing iSWAP oscillations between two spins about 200 micron apart on the same chip~\cite{dijkema2025cavity}. In that experiment, both the spin-spin time evolution and the spin readout took place in the dispersive regime. In  contrast, the observation of vacuum Rabi oscillations poses additional challenges. Notably, for spin readout, either a single microwave photon must be read out directly, or sensing dots must be integrated next to the double dots hosting the spins, or the system must be rapidly pulsed between the resonant and dispersive regime. Moreover, single-qubit operations need to be implemented in the dispersive or uncoupled regime to avoid populating the resonator with photons. Finally, the cavity losses must be as low as possible, which has been a challenge in previous/other spin-photon coupling devices with high-impedance resonators and was solved in this work by using compact and efficient gate filters \cite{@collard2020}.

In this work, utilizing an engineered semiconductor-superconductor hybrid device, we study time-domain interconversion of a single energy excitation between real photons in a superconducting cavity and spin qubits in silicon quantum dots. To observe these vacuum Rabi oscillations, the system was rapidly pulsed between the uncoupled regime, the resonant regime and the dispersive regime. Mastering this control for spins at two ends of the same resonator, we next test the coherent transfer of a single excitation from one spin to the cavity and on to the second spin.
Finally, we verify the Fock-state nature of the photon generated by one qubit, using the other qubit as an analyzer of the photon number by measuring the acceleration in its vacuum Rabi oscillation~\cite{hofheinz2008generation}.\par

\section{Hybrid Device}

A detailed description of the device can be found in Ref.~\cite{@collard2022}. Here, we briefly summarize a few key points. The device is fabricated on an isotopically enriched $^{28}$Si/SiGe heterostructure which is grown by reduced-pressure chemical vapor deposition (\qty{800}{ppm} residual $^{29}$Si in the quantum well). At the two far ends of the resonator stand two double quantum dots (DQDs) that are separated by \qty{250}{\mu m} and host the two spin qubits (Q1 and Q2), respectively (\cref{fig:device}d, e). The aluminum electrodes patterned on top of the substrate are biased with appropriately chosen voltages such that a single electron is captured in each DQD, where it is allowed to hop between the two adjacent dots. A superconducting resonator etched from a thin NbTiN film of \qtyrange{5}{7}{nm} is galvanically connected to one of the two plunger gates of each DQD. Therefore, the electrons are capacitively coupled to the half-wavelength mode of the electric field confined in the resonator~\cite{@mi2017}. The high kinetic inductance of the NbTiN thin film and the nanowire geometry of the resonator enhance the charge-photon coupling strength~\cite{@samkharadze2016}. The bare resonator decay rate is 1.8 MHz.

\begin{figure*}[htbp] 
\center{\includegraphics[width=\linewidth]{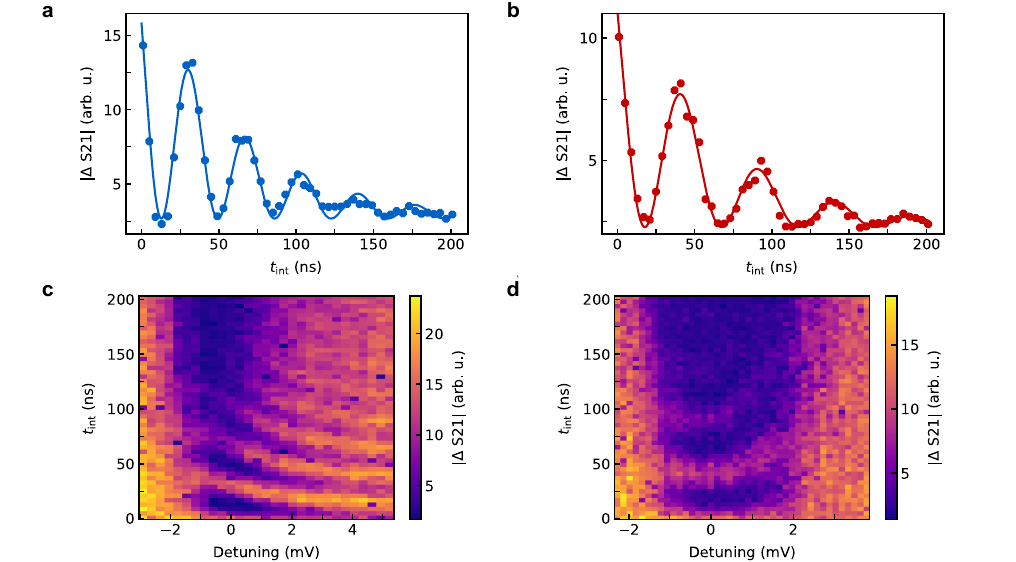}}
\caption{Vacuum Rabi oscillations. \textbf{a,b.} Vacuum Rabi oscillations of (\textbf{a}) Q1 and (\textbf{b}) Q2 . The data points show the magnitude of the measured transmission signal through the resonator as a function of the interaction time $t_{\rm{int}}$ between the resonator and the corresponding qubit. Solid lines are fits to the data using a numerical model of the spin-photon system, including coupling to its environment (see \cref{appendix:simulations} for details). A vacuum Rabi frequency of 27.3 MHz (20.1 MHz) is extracted for Q1 (Q2), which matches spectroscopic vacuum Rabi splittings in early experiments on this device. The data points are averaged over $10^6$ shots. \textbf{c,d.} Vacuum Rabi oscillations measured as a function of the interdot detuning, which is controlled by the voltage pulse on RP. At a detuning voltage of 0~mV, which approximates zero interdot detuning, the oscillations reflect resonant spin-photon exchange, as plotted in \textbf{a,b}. At non-zero detuning, the oscillation speed is affected by both a change in the qubit frequency and a reduction in the electric dipole. The data points are averaged over $10^5$ shots.}

\label{fig:VacuumRabi}
\end{figure*}

When the electron wavefunction is delocalized over the left and the right dot, the so-called ``flopping-mode'' regime~\cite{@benito2019b}, its charge dipole moment is maximized~\cite{@hu2012, @benito2019b}, enabling a strong charge-photon coupling of \qty{192}{MHz} (\cref{fig:device}c). A pair of micromagnets made of cobalt is placed on top of each DQD (\cref{fig:device}d). An external magnetic field polarizes the micromagnets, producing a transverse magnetic field gradient that is essential to both single-spin operations by electric-dipole spin resonance (EDSR)~\cite{pioro2008electrically, kawakami2014electrical} and strong spin-photon coupling through synthetic spin-charge hybridization~\cite{@samkharadze2018, @mi2018}. This hybridization also results in comparatively short spin relaxation rate of 2.7-2.8 MHz (\cref{appendix:simulations}). The two pairs of micromagnets are tilted by 30 degrees relative to each other~\cite{@borjans2019, @collard2022}. When the external magnetic field is set to \qty{53.4}{mT} and is oriented at an angle of \qty{9.42}{degrees} with respect to the interdot axis, the two spin qubits and the cavity photons share the same frequency, $\approx 6.904$ GHz, which is required to implement vacuum Rabi oscillations (\cref{fig:device}b). When there are $(n + 1) \ge 1$ total excitations in the spin-photon system, where $n$ is the photon number when the spin is excited, the spin state and photon number are no longer independent of each other, but form hybridized eigenstates $(\ket{n}\ket{\uparrow}\pm\ket{n+1}\ket{\downarrow})/\sqrt2$ which differ in energy by $2\sqrt n\hbar g_s$, with $g_s$ the spin-photon coupling strength.

By sending voltage pulses to the electrodes using an arbitrary waveform generator (AWG), the electron in each DQD can be rapidly switched between the flopping-mode regime and a configuration where the electron is tightly confined in a single dot. In the latter configuration, the spin-photon coupling is effectively switched off, and this is where we perform single-qubit manipulations and readout (explained later). To circumvent an electrostatic drift in the sample, a fast calibration procedure is implemented as explained in the supplementary information in ref.~\cite{dijkema2025cavity}.

\begin{figure*}[htbp] 
\center{\includegraphics[width=\linewidth]{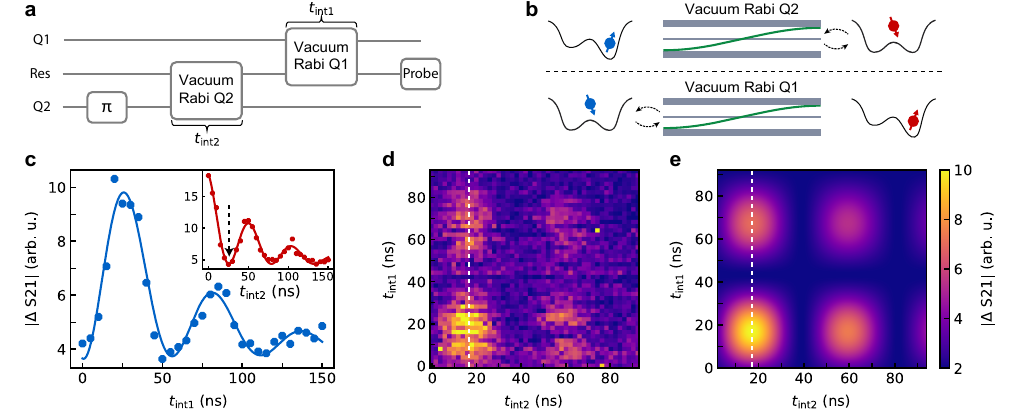}}
\caption{Transfer of excitations between distant spins. \textbf{a.}~Circuit schematic implemented in the experiment. \textbf{b.}~Schematics showing the double dot configuration leading to vacuum Rabi oscillations for Q2 and Q1 respectively. \textbf{c.}~With the resonator being populated and Q1 in its ground state, a vacuum Rabi oscillation with inverted phase is observed. The oscillation frequency is fitted to be 17.8 MHz. Inset, vacuum Rabi oscillation of Q2 used to prepare a photon in the resonator. The dashed arrow indicates the duration of the pulse ($t_{\rm{int2}} = 27$ ns) used to measure the blue data points in the main figure. The data points are averaged over $5\times10^5$ shots. \textbf{d-e.}~Measured and simulated transmission signals as a function of the interaction times for two consecutive vacuum Rabi oscillations. The data points in \textbf{d} are averaged over $10^5$ shots. The white dashed lines correspond to the results plotted in blue in \textbf{c}.}
\label{fig:StateTransfer}
\end{figure*}

\section{Vacuum Rabi Oscillations}

We first probe the vacuum Rabi oscillation between each spin qubit and the cavity. We start with the electron localized in the right dot of one DQD and the cavity empty ($\ket{n=0}$). The spin is initialized to $\ket{\downarrow}$ through spin relaxation, which is very fast, as discussed below. Then a calibrated microwave burst is applied to LP to flip the spin from $\ket{\downarrow}$ to $\ket{\uparrow}$. We next turn on the spin-photon interaction by bringing the DQD near the degeneracy between the (1,0) and (0,1) charge configurations using a pulse on RP with a \qty{5}{ns} ramp. The other qubit remains detuned in frequency from the cavity throughout this experiment. The ramp used to activate the vacuum Rabi oscillation must ensure adiabaticity with respect to the spin-charge hybridization but diabaticity compared to the spin-photon vacuum Rabi splitting ($\hbar2g_s$). We allow the spin and photon to interact for a time $t_{\rm{int}}$, which is expected to lead to an oscillation of the form $\cos(g_st_{\rm{int}})\ket{0}\ket{\uparrow}-i\sin(g_st_{\rm{int}})\ket{1}\ket{\downarrow}$. After a time $t_{\rm{int}}=\pi/2g_s$, a quantum of energy is exchanged between spin and photon~\cite{sillanpaa2007coherent}.

We record the spin population through a dispersive measurement, whereby the resonator frequency is slightly shifted depending on the spin state. To reach the dispersive regime, we bring the electron mildly away from the interdot degeneracy point. In this setting, the spin frequency is slightly detuned from the cavity frequency (by 43 MHz and 35 MHz for Q1 and Q2, respectively) because of the longitudinal magnetic field gradient provided by the micromagnet, as well as by reduced spin-charge hybridization. The transmission of a probe tone is then used as a measure of the spin population (see \cref{appendix:readout})~\cite{@mi2018, dijkema2025cavity}.\par

The experimentally measured cavity transmission as a function of the spin-photon interaction time is plotted in \cref{fig:VacuumRabi}a,b. We observe that the transmission exhibits a decaying sinusoidal oscillation, as expected, from which we extract $2g_s$. Multiple periods of oscillation are visible, corresponding to the repeated exchange of energy between the spin and the photon. The observed initial phase of the oscillation can be attributed to the spin-photon interaction during the detuning ramps (see \cref{supsec:ramps}). In~\cref{fig:VacuumRabi}c,d, we vary both the interaction time and the RP pulse amplitude, with the latter controlling the interdot detuning. As the electron moves away from the degeneracy point, the spin-photon frequency detuning $\Delta$, introduced by smaller spin-charge mixing and the longitudinal magnetic field gradient, is supposed to accelerate the spin-photon oscillation if the spin-photon coupling strength remains unchanged. However, at the same time the reduced charge dipole reduces the charge-photon coupling strength and thus the spin-charge hybridization and spin-photon coupling strength. Indeed, throughout ~\cref{fig:VacuumRabi}d and in~\cref{fig:VacuumRabi}c for negative detuning, we observe slower oscillations away from the charge degeneracy point. An exception occurs at positive voltage detuning for DQD1, where the oscillation accelerates instead, see the right side of~\cref{fig:VacuumRabi}c. Earlier experiments on the same device~\cite{@collard2022} found that DQD1 is subject to a much steeper longitudinal magnetic field gradient than DQD2, and this effect dominates over the suppressed spin-photon coupling strength. We have qualitatively reproduced this asymmetry using numerical simulations in \cref{supsec:VROassymetry}. 
\par

\section{Spin Excitation Transfer}

Vacuum Rabi oscillations can be harnessed to transfer an excitation of one spin to another spin at a distance, a protocol which has utility in quantum networks. When the interaction time is precisely calibrated to produce a $\pi$ rotation, the spin state is coherently exchanged with the photon state in the Fock basis, $\ket{\downarrow} \leftrightarrow \ket{n=0}$ and $\ket{\uparrow} \leftrightarrow \ket{n=1}$. Thus, when the cavity is initially empty, an arbitrary initial state of the spin is transferred to the photon state and the spin is left behind in the ground state $\ket{\downarrow}$. At this point, the qubit state encoded in the photon  can be transferred to a second spin at the other end of the cavity, as we will test below.

\begin{figure}[htbp] 
\center{\includegraphics[width=\linewidth]{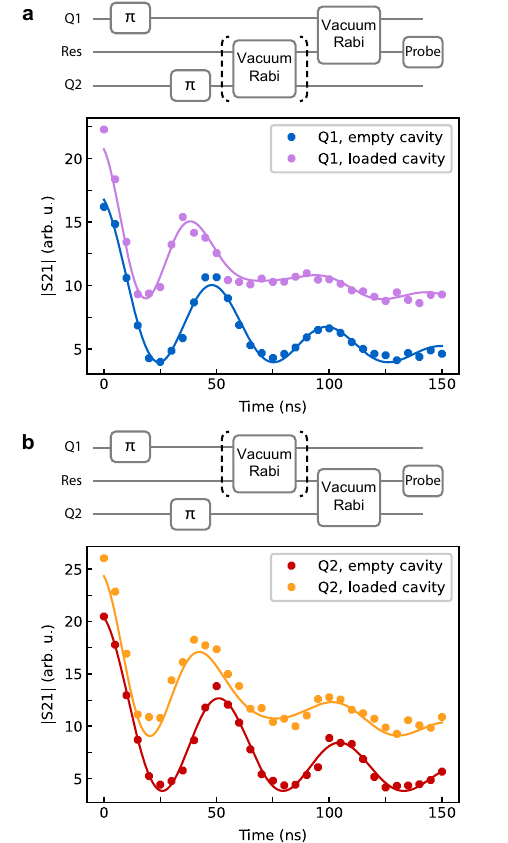}}
\caption{Accelerated vacuum Rabi oscillations. \textbf{a.}~The top panel shows the circuit schematic. Both qubits start in their excited state. Then the spin excitation in Q2 is completely swapped into the resonator and subsequently, the vacuum Rabi oscillation between Q1 and the photon-occupied resonator is measured, as shown in the bottom panel. The vacuum Rabi oscillation is found to be faster by a factor of 1.28 (purple data points and lines) compared to the case of an empty cavity (blue data points and lines), which corresponds to an initial photon number of 0.65. \textbf{b.}~The same experiment is performed with the roles of the two qubits interchanged. Here the vacuum Rabi oscillation is accelerated by a factor of 1.24 which corresponds to an initial photon number of 0.55. The data points are averaged over $10^6$ shots and the solid colored lines represent fits using a numerical model (see \cref{appendix:simulations}).}
\label{fig:Fock}
\end{figure}

The sequence for quantum state transfer is presented in~\cref{fig:StateTransfer}a,b. The first half of the sequence is identical to the vacuum Rabi sequence as mentioned previously, only with the interaction time fixed to $t_{\rm{int}} = \pi/2g_s \approx 27$~ns, indicated by the dashed arrow in the inset of ~\cref{fig:StateTransfer}c, such that the state of Q2 is mapped onto the photon state  (in quantum gate language, the operation corresponds to an iSWAP operation in the one-excitation subspace, up to a phase correction). Subsequently, Q1, which starts off in $\ket{\downarrow}$, is allowed to interact with the cavity that has been populated with a photon. As shown in ~\cref{fig:StateTransfer}c, this results in a vacuum Rabi oscillation with inverted phase compared to the outcome in~\cref{fig:VacuumRabi}. When also the second vacuum Rabi oscillation is halted after half a period, an excitation is transferred from Q2 to Q1 through the photon. Noticeably, such a state transfer scheme only works when the cavity and the ``receiver'' qubit start off in their ground states.\par

To further test the coherence of the state transfer process, we simultaneously sweep the interaction time of both vacuum Rabi oscillations, starting from Q2 in $\ket{\downarrow}$, Q1 in $\ket{\uparrow}$ and the cavity in $\ket{0}$. We observe a series of periodically appearing circular patterns (\cref{fig:StateTransfer}d), similar to those reported in Ref.~\cite{sillanpaa2007coherent} for superconducting qubits. When $t_{\rm{int2}} = \pi/2g_s$ (white dashed vertical lines in \cref{fig:StateTransfer}d,e), the excitation from Q2 is fully mapped to the photon and we see the inverted vacuum Rabi oscillation of Q1 as a function of $t_{\rm{int1}}$ in~\cref{fig:StateTransfer}d, similar to that of ~\cref{fig:StateTransfer}c. When $t_{\rm{int2}}$ is taken half as long, the photon and Q2 become entangled. If we next map the photon state onto the state of Q1 using $t_{\rm{int1}} \approx 56$ ns, the two spins, separated by about 250 $\mu$m, are expected to become entangled~\cite{nikitchenko2026photon}. Whereas the oscillations in ~\cref{fig:StateTransfer}d support this interpretation, quantum state tomography would be needed to firmly demonstrate the creation of an entangled state.  Numerical simulations of the dynamics of the two spins interacting with the resonator, including effects of decoherence, are in good agreement with the data, compare~\cref{fig:StateTransfer}e (see \cref{appendix:simulations}).\par


\section{Fock state generation of photons}

We now further analyze the cavity state after transferring a single excitation from a quantum dot spin, and verify that the photon is in a Fock state after half a period of vacuum Rabi oscillations. According to the Jaynes-Cummings model, when there are in total $n$ excitations in the spin-photon hybrid system, the frequency of vacuum Rabi oscillations is scaled by a factor of $\sqrt{n}$~\cite{hofheinz2008generation}. To measure the accelerated vacuum Rabi oscillation, we initialize both qubits into $\ket{\uparrow}$ with the cavity empty, such that two excitations are present in the system. Then we prepare the cavity with ideally exactly one photon by swapping an excitation from one of the spins with the photon. Next we probe the vacuum Rabi oscillation between the populated cavity and the other spin. If there is no loss during the process, the vacuum Rabi frequency should be faster by a factor of $\sqrt{2}$ compared to the case with only one total excitation. In practice, however, we expect two effects to lead to deviations from this ideal scenario. First, energy loss and dephasing during the first vacuum Rabi oscillation will prepare the cavity with an average photon number of less than 1, which results in an initial scaling factor smaller than $\sqrt{2}$. Second, the scaling factor is expected to decrease exponentially as a function of the interaction time during the second vacuum Rabi oscillation, because the excess excitation will be damped. Therefore, we first fit the simulated time evolution to the oscillation with initial condition of an empty resonator.
Subsequently, we estimate the loaded initial resonator state $(1-p)\ket{0}\bra{0} + p\ket{1}\bra{1}$ by varying $p$ until the time evolution matches the measured accelerated oscillation (see \cref{appendix:simulations} for details).

We first implement the experiment using Q2 to prepare the cavity and Q1 to measure the accelerated oscillation, and then perform the same experiment with the qubit roles interchanged. The data is shown in \cref{fig:Fock}a and \cref{fig:Fock}b respectively. For comparison, we show in the same panels Rabi oscillations with the cavity initially empty. As expected, the Rabi oscillation is accelerated initially for the case with a populated cavity, progressively slowing down as photons are lost. This behavior is well reproduced by the master equation simulation, shown as solid lines. We extract an initial photon number of 0.65 when using Q2 to prepare the cavity and 0.55 when using Q1. As a cross-check, the fitted simulation of the regular vacuum Rabi oscillation, which is used to prepare the cavity, yields initial photon numbers of 0.69 and 0.67 at the start of the second Rabi oscillation. In future devices with improved lifetimes of hybrid spin-photon systems, the same operation to prepare a photon Fock state can be repeated to reach higher photon numbers~\cite{hofheinz2008generation, @wang2008}.

\section{Conclusion}

In conclusion, we report the transfer of an energy excitation between two single spins and a microwave photon. Vacuum Rabi oscillations are observed in the time domain between each spin and the cavity photon separately. Concatenating two vacuum Rabi oscillations involving two different spins, we can detect the vacuum Rabi oscillation of one spin by monitoring the state of the other spin. When introducing two excitations into the system at the start, we can prepare a photon Fock state and obtain accelerated vacuum Rabi oscillations. \par

Several periods of vacuum Rabi oscillations can be seen even though this particular device suffers from broader spin linewidths and faster relaxation rates than typical in silicon quantum dots. This is true both when measured at the charge degeneracy point and with the spin confined in a single dot. We presume this is caused by excessive background charge noise, of an origin that is not clear, which couples to the spin through the magnetic field gradient from the micromagnet. It is interesting to compare the quality of the vacuum Rabi oscillations with that of the iSWAP oscillations taken in the dispersive regime on the same device~\cite{dijkema2025cavity}. For the present device, given the short-lived spin coherence, the reduced sensitivity to photon loss in the dispersive regime does not compensate for the slower oscillations by a factor of $g_s/\Delta$. \par

Future devices with state-of-the-art charge noise and spin linewidths, should enable high-fidelity coherent quantum state transfer, entanglement of distant spins, and Fock-state generation with higher photon numbers. These ingredients will allow modular spin qubit architectures and, with the addition of quantum transducers, interfaces of spin qubits to photonic quantum networks. Additionally, the capability to prepare photons in a cavity opens up the possibility to study quantum algorithms and hybrid quantum simulations that involve both bosonic and fermionic degrees of freedom in the same system. 

\section*{Acknowledgements}
The authors thank W. Oliver for providing the TWPA, L. P. Kouwenhoven and his team for access to the NbTiN film deposition, F. Alanis Carrasco for assistance with sample fabrication, L. DiCarlo and his team for access to the $^3$He cryogenic system, O. Benningshof, R. Schouten and R. Vermeulen for technical assistance, and other members of the spin-qubit team at QuTech for useful discussions. This research was supported by the Dutch Research Council (NWO) via the National Growth Fund program Quantum Delta NL (grant no. NGF.1623.23.024) and the European Union’s Horizon 2020 research and innovation programme under the Grant Agreement No. 951852 (QLSI project), the European Research Council (ERC Synergy Quantum Computer Lab), the Dutch Ministry for Economic Affairs through the allowance for Topconsortia for Knowledge and Innovation (TKI), and the Netherlands Organization for Scientific Research (NWO/OCW) as part of the Frontiers of Nanoscience (NanoFront) program.\par

\textbf{Data and code availability} Data supporting this work and codes used for data processing are available at open data repository 4TU~\cite{Data}.


\textbf{Competing interests}
The authors declare no competing interests.

\bibliography{library, library_2}

\appendix
\onecolumngrid

\section{Simulations}
\label{appendix:simulations}
As schematically seen in Figure 1 of the main text, the system under study contains three relevant degrees of freedom: the photons inside the resonator, and the spin and position of one electron in a double quantum dot (DQD).
In this work, only a single DQD is tuned to the spin-photon interaction point at each time, while the other is parked away from zero detuning where the interaction is negligible. 
The system Hamiltonian can be written as
\begin{equation}
\label{eq:dqdhamiltonian}
H = H_{\rm res} + H_{\rm DQD} + H_{\rm int}.
\end{equation}
Here $H_{\rm res}= \hbar\omega_r a^\dagger a$ describes the fundamental mode of the resonator, where $a^\dagger$ and $a$ are the photon creation and annihilation operators. 
Each interacting DQD is operated in the so-called flopping-mode regime \cite{@benito2019b, @croot2020}, where an electron is delocalized over the double quantum dot under the effect of a spatially varying magnetic field.
The Hamiltonian for each DQD can therefore be written as \cite{dijkema2025cavity, @benito2019}
\begin{equation}
H_{\rm DQD} = \frac{1}{2}\left(\varepsilon\tau_{z}+2t_{c}\tau_{x}+g_e\mu_B (\boldsymbol{B}+\boldsymbol{\Delta B}\tau_{z}/2)\cdot \boldsymbol{\sigma}\right),
\end{equation}
with $\boldsymbol{\sigma} = (\sigma_x, \sigma_y, \sigma_z)^T$. Here $\tau_\alpha$ and $\sigma_\alpha$ ($\alpha = x, y, z$) are the Pauli operators for the electron's position (left, right) and spin ($\uparrow, \downarrow$) respectively, $g_e\approx2$ is the Landé g-factor in silicon, and $\mu_B$ is the Bohr magneton.
Near zero interdot detuning ($\varepsilon=0$), the electron wavefunction becomes delocalized over the two dots due to the tunnel coupling $t_c$, and the charge eigenstates are the bonding and antibonding orbital states $\ket{\pm}$, which are split in energy by $\Omega = \sqrt{\varepsilon^2+4t_c^2}$.
The average magnetic field $\boldsymbol{B} = (0,0,B_z)^T$ is taken to point along the $z$-axis.
The magnetic field difference between the two quantum dots  $\boldsymbol{\Delta B} = (\Delta B_x, 0, \Delta B_z)^T$ includes both a transversal component $\Delta B_x$, which leads to hybridization of spin and charge states, and a longitudinal component $\Delta B_z$, which causes the spin frequency in each of the dots to differ \cite{@beaudoin2016}. \par

The coupling between the resonator and the DQD charge degree of freedom is described by a standard quantum Rabi model
\begin{equation}
H_{\rm int} = \hbar g_c (a+a^\dagger)\tau_z,
\end{equation}
where $g_c$ is the charge-photon coupling strength.
This charge-photon coupling in combination with the spin-charge hybridization near zero detuning gives rise to an effective spin-photon coupling $g_s$~\cite{@beaudoin2016, @benito2017}. \par

\subsection{Effective spin-photon model}
\label{supsec:effectivemodel}
To enable iterative fitting of the simulations to the experimental data, we eliminate the charge degree of freedom in the limit $\max(\hbar g_c, g\mu_B \Delta B_x)\ll 2t_c$ and set $\varepsilon=0$, $\Delta B_z=0$ \cite{dijkema2025cavity}.
The effect of nonzero $\varepsilon$ and $\Delta B_z$ will be investigated in \cref{supsec:VROassymetry}.
We then apply the rotating wave approximation to arrive at an effective spin-photon model, described by the Jaynes-Cummings Hamiltonian
\begin{equation}
H_{JC} = \hbar\omega^*_r a^\dagger a + \frac{\hbar\omega_q}{2}\sigma_z + \hbar g_s(a^\dagger \sigma_-+a\sigma_+).
\end{equation}
Importantly, eliminating the charge degree of freedom neglects the relevant resonance frequency shift that originates from the dispersive charge-photon interaction.
Therefore, instead of the bare resonator frequency $\omega_r$, we use the loaded resonator frequency $\omega^*_r=\omega_r-\chi_c$, where 
\begin{equation}
\chi_c = \hbar^2 g_c^2 \left(\frac{1}{\Omega-\hbar\omega_r} + \frac{1}{\Omega+\hbar\omega_r}\right)
\end{equation}
is the charge dispersive shift. 
We remark that the large charge-photon detuning $\Omega-\hbar\omega_r$ in our experiments causes both co-rotating and counter-rotating (Bloch-Siegert shift) terms to contribute substantially to the shift \cite{kohlerDispersiveReadoutUniversal2018}.
The ``qubit'' frequency $\omega_q$ corresponds to the predominantly spin-like transition from the DQD ground state, for which the effective spin-photon coupling strength can be approximated by
\begin{equation}
    \label{eq:gs}
    g_s \approx g_c \cos\theta \sin\Phi/2,
\end{equation}
with delocalization angle $\theta=\arctan\frac{\varepsilon}{2t_c}$ and spin-charge mixing angle $\Phi=\arctan\frac{g_e\mu_B \Delta B_x \cos\theta}{2(\Omega-g_e\mu_B B_z)}$ \cite{@benito2017}.

\subsection{Numerical simulations and decoherence}
\label{supsec:masterequation}
The system dynamics are simulated using the Lindblad master equation
\begin{equation}
\label{eq:masterequation}
\frac{d\rho}{dt} = -\frac{i}{\hbar}(H \rho-\rho H) + \sum_i \mathcal{D}_i (\rho)
\end{equation}
with Lindblad dissipation terms $\mathcal{D}_i(\rho) = \gamma_i\left(L_i\rho L_i^\dagger - \frac{1}{2}\{L_i^\dagger L_i,\rho\}\right)$.
The simulations are implemented using QuTip~\cite{qutip5}, with a truncated resonator Hilbert space including up to $N_{\rm max} = 5$ photons in the resonator. \par

The initial state of the system depends on the respective experiment and is described by a density matrix of the form $\rho_{\rm init} = \rho_{\rm init, res} \otimes \rho_{\rm init, q}$.
We assume the spin state preparation to be perfect, such that $\rho_{\rm init, q} = \ket{\uparrow}\bra{\uparrow}$ in all simulations except for the swapped vacuum Rabi simulation in \cref{fig:StateTransfer}b, where $\rho_{\rm init, q} = \ket{\downarrow}\bra{\downarrow}$.
For the standard vacuum Rabi oscillation (main text \cref{fig:VacuumRabi}, \cref{fig:StateTransfer}a), the resonator is initially in the vacuum state $\rho_{\rm init, res} = \ket{0}\bra{0}$.
Meanwhile, in the swapped (main text \cref{fig:StateTransfer}c) and accelerated (main text \cref{fig:Fock}) vacuum Rabi experiments of the main text, the resonator is ideally populated by a single photon using an initial vacuum Rabi oscillation.
Since our Fock-state preparation has success probability $p<1$, we use a mixed state $\rho_{\rm init, res} = (1-p)\ket{0}\bra{0} + p\ket{1}\bra{1}$ as the initial resonator state in the respective simulations. \par

\subsection{Fitting procedure}
To fit the experimental data, we perform a master equation simulation using the effective spin-photon Hamiltonian $H_{JC}$, and include the following dissipation terms in \cref{eq:masterequation}
\begin{align}
    \gamma_1 &= \kappa^*\ \ \ L_{1} = a,\\
    \gamma_2 &= \gamma_{1,s}\  \ L_{2} = \sigma_-,\\
    \gamma_3 &= \gamma_{\phi,s}\  \ L_{3} = \sigma_z.
\end{align}
Similar to the resonator frequency, the photon decay rate $\kappa^*$ is modified from its bare value $\kappa$ due to the dispersive interaction with the charge degree of freedom.
Additionally, the model includes spin relaxation at rate $\gamma_{1,s}$ and pure qubit dephasing at rate $\gamma_{\phi,s}$. \par

We compute the spin-up probability as $P_\uparrow = \text{Tr}[\ket{\uparrow}\bra{\uparrow}\rho(t_{\rm sim})]$, which we finally convert to a transmission signal using a linear mapping
\begin{equation}
|S_{21}|_{\rm sim} = aP_\uparrow + b,
\end{equation}
where $a, b$ are fitted parameters.
For the experimental parameters used, the linearization of the readout signal captures well the response from input-output theory \cite{@benito2017, bonsenProbingJaynesCummingsLadder2023}.\par

Some of the measured vacuum Rabi oscillations reveal an initial phase different from zero, which we attribute to the fact that the spin and photon can interact during the ramps from and to zero detuning (see \cref{supsec:ramps}).
This causes the total interaction time to be different from $t_{\rm int}$, which is the interaction time excluding the ramps.
To reproduce this effect, we include an initial delay $t_0$ and use final time $t_{\rm sim}=t_0 + t_{\rm int}$ for simulating the time evolution of the density matrix $\rho(t_{\rm sim})$.
We find that the experimental data can be fitted consistently when using five fitting parameters: the spin-photon coupling strength $g_s$, the spin relaxation rate $\gamma_{1,s}$, the scaling parameter $a$, the signal offset $b$, and the initial delay $t_0$.
Since the observed vacuum Rabi oscillations are mainly limited by relaxation, we neglect pure spin dephasing and fix $\gamma_{\phi,s}=0$.
The loaded resonator frequency $\omega^*_r/2\pi = \SI{6.9040}{\giga\hertz}$ and photon decay rate $\kappa^* = \SI{2.5}{\mega\hertz}$ are extracted from independent measurements with either of the DQDs at zero detuning.
The spin is assumed to be on resonance with the photon,  i.e.,  $\omega_q=\omega^*_r$. \par

The fitting is performed by wrapping the QuTip simulation inside a LMFIT Model \cite{lmfit_1_3_2} that uses the Nelder-Mead method to find an optimal set of parameters (full code available at \cite{Data}).
Each dataset is fitted independently, with the exception of the accelerated vacuum Rabi oscillations in main text~\cref{fig:Fock}.
In that case, we first fit the simulated time evolution to the oscillation with the initial empty resonator state $\rho_{\rm init, res}=\ket{0}\bra{0}$.
Subsequently, we use the fitted parameters to perform the simulation with the initial loaded resonator state $\rho_{\rm init, res} = (1-p)\ket{0}\bra{0} + p\ket{1}\bra{1}$, where we manually vary $p$ until the simulated time evolution matches the experimental results.
The resulting values for $p$ approximately match the simulated photon number after a half-period vacuum Rabi oscillation of the other qubit. \par

The parameters used in the state transfer simulation as a function of $t_{\rm int,1}$ and $t_{\rm int,2}$, shown in main text \cref{fig:StateTransfer}e, are not fitted.
Instead, they are estimated from the peaks in signal intensity ($g_s, t_0$) and the decay ($\gamma_{1,s}$) of the experimental data in \cref{fig:StateTransfer}d.
We remark that the experiment in main text \cref{fig:StateTransfer}d was performed under different tuning conditions than the experiment in main text \cref{fig:StateTransfer}c.
The simulation parameters for each of the datasets presented in the main text are presented in \cref{tab:simulationparams}. \par

\sisetup{separate-uncertainty}
\setlength{\tabcolsep}{8pt}
\renewcommand{\arraystretch}{1.2}
\begin{table}[h]
\caption{Simulation parameters in the effective spin-photon model. The fitted values and standard errors are obtained using the Nelder-Mead fitting method implemented in LMFIT \cite{lmfit_1_3_2}}.
\label{tab:simulationparams}
\begin{tabular}{lllcc}
\hline
\textbf{General}                         & \textbf{Symbol}    & \textbf{Determination} & \multicolumn{2}{c}{\textbf{Value}}            \\ \hline
Dispersively shifted resonator frequency & $\omega^*_r/2\pi$  & Measured               & \multicolumn{2}{c}{\qty{6.9040}{GHz}}                \\
Loaded resonator decay rate              & $\kappa^*_r/2\pi$  & Measured               & \multicolumn{2}{c}{\qty{2.5}{MHz}}                   \\
Unloaded resonator decay rate              & $\kappa_r/2\pi$  & Measured               & \multicolumn{2}{c}{\qty{1.8}{MHz}}                   \\
Maximal included photon number           & $N_{\rm max}$      & Assumed                & \multicolumn{2}{c}{5}                         \\
Qubit frequency                          & $\omega_q/2\pi$    & Assumed                & \multicolumn{2}{c}{\qty{6.9040}{GHz}}                \\
Qubit pure dephasing rate                & $\gamma_{\phi,s}/2\pi$ & Assumed                & \multicolumn{2}{c}{0}                         \\ \hline
\textbf{Single-qubit VRO (Fig. 2)}       & \textbf{}          & \textbf{}              & \textbf{Q1 (Fig. 2a)} & \textbf{Q2 (Fig. 2b)} \\ \hline
Spin-photon coupling strength            & $g_s/2\pi$         & Fitted                 & \qty{13.66\pm0.10}{MHz}              & \qty{10.06\pm0.08}{MHz}              \\
Qubit relaxation rate                    & $\gamma_{1,s}/2\pi$& Fitted                 & \qty{2.68\pm0.26}{MHz}            & \qty{2.80\pm0.22}{MHz}               \\
initial delay                            & $t_0$              & Fitted                 & \qty{5.40\pm0.29}{ns}            & \qty{6.91\pm0.35}{ns}                \\
scaling parameter                        & $a$                & Fitted                 & \qty{1.80\pm0.07e-3}{}           & \qty{1.22\pm0.04e-3}{}                \\
offset                                   & $b$                & Fitted                 & \qty{2.67\pm0.11e-4}{}               & \qty{2.28\pm0.06e-4}{}                \\ \hline
\textbf{State transfer 1D (Fig. 3c)}   & \textbf{}          & \textbf{}                & \textbf{Q1 (Fig. 3c)} & \textbf{Q2 (inset)} \\ \hline
Spin-photon coupling strength            & $g_s/2\pi$         & Fitted                 & \qty{8.90\pm0.11}{MHz}             & \qty{9.40\pm0.14}{MHz}               \\
Qubit relaxation rate                    & $\gamma_{1,s}/2\pi$    & Fitted             & \qty{2.60\pm0.52}{MHz}           & \qty{2.46\pm0.31}{MHz}               \\
initial delay                            & $t_0$              & Fitted                 & \qty{0.0}{ns}\footnote{In this case, the fitting method converged to the imposed lower bound, $t_0=0$, up to numerical precision.}                & \qty{1.03\pm0.63}{ns}                \\
scaling parameter                        & $a$                & Fitted                 & \qty{1.44\pm0.09e-3}{}                & \qty{1.44\pm0.06e-3}{}                \\
offset                                   & $b$                & Fitted                 & \qty{3.69\pm0.14e-4}{}                & \qty{4.36\pm0.16e-4}{ }               \\
initial photon number                    & $p$                & Estimated              & \qty{0.65}{}                  & \qty{0}{}                     \\ \hline
\textbf{State transfer 2D (Fig. 3e)}     & \textbf{}          & \textbf{}              & \textbf{Q1}           & \textbf{Q2}           \\ \hline
Spin-photon coupling strength            & $g_s/2\pi$         & Estimated              & \qty{10.0}{MHz}              & \qty{11.9}{MHz}              \\
Qubit relaxation rate                    & $\gamma_{1,s}/2\pi$& Estimated              & \qty{3.0}{MHz}               & \qty{3.0}{MHz}               \\
initial delay                            & $t_0$              & Estimated              & \qty{7.0}{ns}                & \qty{5.0}{ns}                \\ \hline
\textbf{Accelerated VRO (Fig. 4)}        & \textbf{}          & \textbf{}              & \textbf{Q1 (Fig. 4a)} & \textbf{Q2 (Fig. 4b)} \\ \hline
Spin-photon coupling strength            & $g_s/2\pi$         & Fitted                 & \qty{9.83\pm0.12}{MHz}             & \qty{9.34\pm0.09}{MHz}               \\
Qubit relaxation rate                    & $\gamma_{1,s}/2\pi$& Fitted                 & \qty{2.37\pm0.25}{MHz}             & \qty{1.37\pm0.22}{MHz}               \\
initial delay                            & $t_0$              & Fitted                 & \qty{1.02\pm0.52}{ns}                & \qty{1.34\pm0.48}{ns}                \\
scaling parameter                        & $a$                & Fitted                 & \qty{1.30\pm0.04e-3}{}                & \qty{1.68\pm0.05e-3}{}            \\
offset                                   & $b$                & Fitted                 & \qty{3.97\pm0.13e-4}{}                & \qty{3.83\pm0.16e-4}{}                \\
initial photon number                    & $p$                & Estimated              & \qty{0.65}{}                  & \qty{0.55}{}                  \\ \hline
\end{tabular}
\end{table}

\subsection{Asymmetry versus interdot detuning}
\label{supsec:VROassymetry}
In this section, we numerically reproduce the observed asymmetry in the vacuum Rabi oscillations versus interdot detuning seen in main text \cref{fig:VacuumRabi}c,d.
As discussed in the main text, we expect this asymmetry to originate from a difference in magnetic field gradient between both DQDs.
To capture this effect, we simulate the system using the full DQD Hamiltonian in \cref{eq:dqdhamiltonian}.
We first transform the Hamiltonian to the hybridized eigenbasis of charge states
\begin{equation}
\label{eq:Hdqd_chargebasis}
    \tilde{H}_{\rm DQD} = \frac{\Omega}{2} \tilde{\tau}_z + \frac{g_e\mu_B}{2} \left[B_z\sigma_z + (\sin\theta\tilde{\tau}_z - \cos\theta\tilde{\tau}_x)(\Delta B_x\sigma_x + \Delta B_z\sigma_z)/2 \right],
\end{equation}
where $\Omega=\sqrt{\varepsilon^2+4t_c^2}$ and $\theta=\arctan\frac{\varepsilon}{2t_c}$ \cite{@benito2019b}.
Here $\tilde{\tau}_\alpha$ ($\alpha = x, y, z$) are the Pauli operators in the charge eigenbasis of bonding and antibonding orbitals $\ket{\pm}$.
In this basis, the charge-photon interaction Hamiltonian becomes
\begin{equation}
\label{eq:Hint_chargebasis}
    \tilde{H}_{\rm int} = \hbar g_c (a+a^\dagger)(\sin\theta\tilde{\tau}_z - \cos\theta\tilde{\tau}_x) \,.
\end{equation}
The system Hamiltonian $\tilde{H} = H_{\rm res}+\tilde{H}_{\rm DQD}+\tilde{H}_{\rm int}$ is then used to perform the master equation simulation described in \cref{supsec:masterequation}, including the following Lindblad dissipation terms
\begin{align}
    \gamma_1 &= \kappa\ \ \ \ \ L_{1} = a,\\
    \gamma_2 &= \gamma_{1,c}\  \ L_{2} = \tilde{\tau}_-.
\end{align}
Performing this simulation for varying interdot detuning $\varepsilon$, the aim is to find magnetic field components $B_z, \Delta B_x, \Delta B_z$ for each DQD such that the simulation qualitatively reproduces the experimentally observed behavior in main text \cref{fig:VacuumRabi}c,d, without aiming for full quantitative agreement.
The remaining Hamiltonian parameters $\omega_r$, $t_c$, $g_c$ and resonator decay rate $\kappa$ are determined from spectroscopic measurements.
Finally, the charge relaxation rate $\gamma_{1,c}$ is chosen such that the visibility of the simulated vacuum Rabi oscillations roughly matches that of the experimental data.  \par

\begin{figure}[htbp] 
\center{\includegraphics[width=\textwidth]{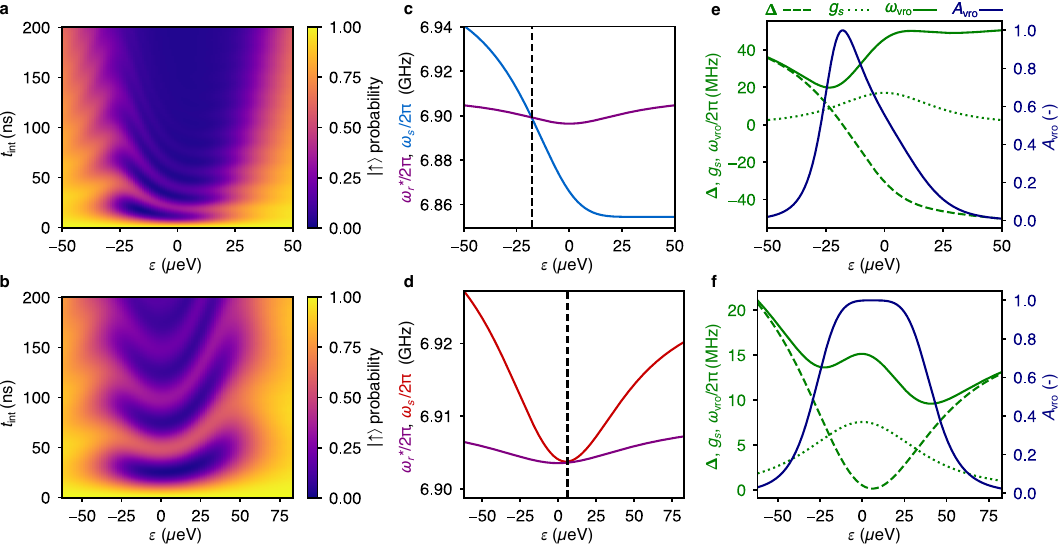}}
\caption{Asymmetry in the vacuum Rabi oscillations versus interdot detuning. \textbf{a,b.}~Simulated spin-up probability as a function of interdot detuning $\varepsilon$ and interaction time $t_{\rm int}$.
In these simulations, we use $\omega_r/2\pi=$~\qty{6.9105}{GHz}, $\kappa/2\pi=$~\qty{1.8}{MHz}, $g_c/2\pi=$~\qty{192}{MHz}, $\gamma_{1,c}/2\pi=$~\qty{1}{GHz}, and $\Delta B_x=$~
\qty{40}{mT}. The remaining simulation parameters are different for DQD1 (a): $2t_c/h=$~\qty{10}{GHz}, $B_z=$~\qty{246.0}{mT}, $\Delta B_z=$~\qty{4}{mT}, and DQD2 (b): $2t_c/h=$~\qty{14}{GHz}, $B_z=$~\qty{246.9}{mT}, $\Delta B_z=$~\qty{0.5}{mT}. \textbf{c,d.}~Effective spin and resonator frequencies as a function detuning for DQD1 (c) and DQD2 (d). The effective spin frequency is calculated as $\omega_{s}/2\pi = (E_1-E_0)/h$, where $E_{i}$ are the eigenenergies of $\tilde{H}_{\rm DQD}$. The effective resonator frequency is calculated as $\omega^*_r/2\pi = (\omega_r-\chi_c)/2\pi$. The dashed lines indicate the interdot detuning at which the spin-photon detuning $\Delta = \omega_s-\omega^*_r$ is minimized. \textbf{e,f.}~Estimation of the frequency and amplitude (solid lines) of the vacuum Rabi oscillations for DQD1 (e) and DQD2 (f) based on the effective spin-photon detuning $\Delta$ (dashed line) and the effective spin-photon coupling strength $g_s$ (dotted line).
}
\label{supfig:vrosim_vs_detuning}
\end{figure}

The simulated spin-up probability maps in \cref{supfig:vrosim_vs_detuning}a,b qualitatively reproduce the observed asymmetry of the interdot detuning (see main text \cref{fig:VacuumRabi}c,d).
As discussed in the main text, moving away from $\varepsilon=0$ affects the vacuum Rabi oscillation in two ways: (1) the effective spin-photon detuning changes due to a combination of the longitudinal gradient $\Delta B_z$ and a weaker spin-charge hybridization, and (2) the effective spin-photon coupling is reduced due to the smaller charge dipole.
The first effect is elucidated in \cref{supfig:vrosim_vs_detuning}c,d, which show the effective spin ($\omega_s$) and resonator ($\omega^*_r$) frequencies.
For DQD1, the stronger longitudinal gradient ($\Delta B_z$) causes the spin frequency to drop below the resonator frequency for $\varepsilon>$~\qty{-18}{\micro eV}, while for DQD2 the spin frequency is minimized near $\varepsilon=$~\qty{6}{\micro eV}.
The second effect, a reduction in $g_s$ away from $\varepsilon=0$, is captured by Eq.~\eqref{eq:gs}.
The interplay of these two effects for each of the DQDs is illustrated in \cref{supfig:vrosim_vs_detuning}e,f, which show the estimated vacuum Rabi oscillation frequency
\begin{equation}
    \omega_{\rm vro} = \sqrt{\Delta^2+4g_s^2},
\end{equation}
and amplitude
\begin{equation}
    A_{\rm vro} = \frac{4g_s^2}{\Delta^2+4g_s^2},
\end{equation} 
which are calculated from the effective spin-photon detuning $\Delta = \omega_s-\omega^*_r$ and coupling strength $g_s$ in \cref{eq:gs}.
For DQD1, the rapidly increasing spin-photon detuning leads to the observed speedup of the vacuum Rabi oscillations for increasing $\varepsilon$.
Increasing $\varepsilon$ further, the oscillation is damped by the decreasing spin-photon coupling strength, as signified by the decrease in $A_{\rm vro}$.
For DQD2, the spin-photon detuning remains positive and is smaller than for DQD1.
In the range where the oscillations are most visible ($A_{\rm vro}>0.5$), the reduction in spin-photon coupling strength dominates over the increase in spin-photon detuning, leading to the observed slowdown of the oscillations away from $\Delta=0$ in this case. \par

\subsection{Phase accumulation during the detuning ramps}
\label{supsec:ramps}
Finally, we investigate the origin of the observed initial phase at $t_{\rm int}=0$ of the vacuum Rabi oscillations, which was accounted for in the effective spin-photon model (\cref{supsec:effectivemodel}) by including an initial delay $t_0$ as a fitting parameter.
As discussed in the main text, we attribute this initial phase to the spin-photon interaction during the detuning ramps.
Here, we simulate this effect using the model from \cref{supsec:VROassymetry}, where we now make the interdot detuning explicitly time-dependent, i.e., $\varepsilon=\varepsilon(t)$.
Inserting $\varepsilon(t)$ into \cref{eq:Hdqd_chargebasis} and \cref{eq:Hint_chargebasis} leads to a time-dependent Hamiltonian $\tilde{H}(t)$.
We solve the respective time-dependent master equation numerically using QuTiP~\cite{qutip5}. \par

\begin{figure}[htbp] 
\center{\includegraphics[width=\textwidth]{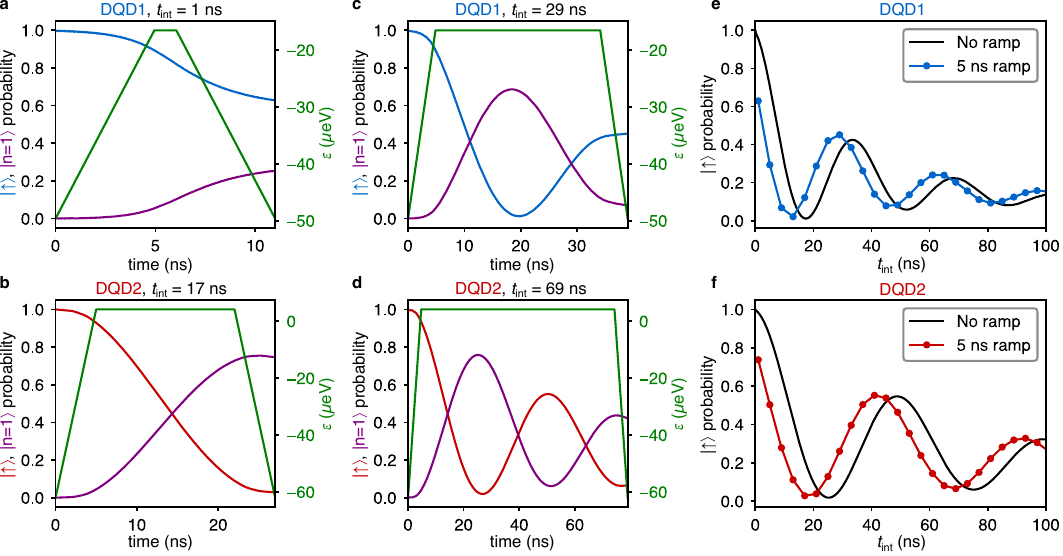}}
\caption{Detuning ramp simulations. \textbf{a-d}~Simulated spin ($\ket{\uparrow}$) and photon ($\ket{n=1}$) excitation probabilities for several interaction times $t_{\rm int}$, throughout the shown interdot detuning profiles $\varepsilon(t)$. The simulation parameters are the same as in \cref{supfig:vrosim_vs_detuning}: $\omega_r/2\pi=$~\qty{6.9105}{GHz}, $\kappa/2\pi=$~\qty{1.8}{MHz}, $g_c/2\pi=$~\qty{192}{MHz}, $\gamma_{1,c}/2\pi=$~\qty{1}{GHz}, and $\Delta B_x=$~
\qty{40}{mT}; for DQD1 (a,c): $2t_c/h=$~\qty{10}{GHz}, $B_z=$~\qty{246.0}{mT}, $\Delta B_z=$~\qty{4}{mT}; for DQD2 (b,d): $2t_c/h=$~\qty{14}{GHz}, $B_z=$~\qty{246.9}{mT}, $\Delta B_z=$~\qty{0.5}{mT}.
Initially, $\varepsilon(t)$ is at the most negative detuning in panels \cref{supfig:vrosim_vs_detuning}a,b, which corresponds to the readout point in the experiment (see \cref{appendix:readout}).
A \qty{5}{ns} ramp then takes $\varepsilon(t)$ to the point where spin-photon resonance occurs (see \cref{supfig:vrosim_vs_detuning}c,d), where it remains for the interaction time $t_{\rm int}$, after which it is ramped back to the readout point.
\textbf{e,f}~Vacuum Rabi oscillations for DQD1~(e) and DQD2~(f), obtained by repeating the detuning ramp simulation for the shown range of interaction times.
Comparing to a simulation where $\varepsilon$ is fixed at the interaction point (``No ramp''), the oscillations for DQD1 (DQD2) appear shifted by \qty{5}{ns}~(\qty{7}{ns}).
}
\label{supfig:vrosim_ramps}
\end{figure}

The simulation results in \cref{supfig:vrosim_ramps} demonstrate that the spin-photon interaction during the detuning ramps can drastically alter the total spin-photon interaction time.
Neglecting interactions during the ramps, the expected final spin-up probability is $P_\uparrow = \cos^2 (g_s t_{\rm int})$, which yields for $t_{\rm int}=$~\qty{1}{ns} and $g_s/2\pi>$~\qty{10}{MHz} a spin-up probability of $P_\uparrow>0.99$.
However, by including the interactions during the ramps, our simulation yields $P_\uparrow\approx 0.63$ for $t_{\rm int}=$~\qty{1}{ns} (\cref{supfig:vrosim_ramps}a), exemplifying the prolonged spin-photon interaction time.
Performing this simulation for a range of interaction times, we find that the vacuum Rabi oscillations for DQD1 (\cref{supfig:vrosim_ramps}e) and DQD2 (\cref{supfig:vrosim_ramps}f) appear shifted in time by \qty{5}{ns} and \qty{7}{ns}, respectively.
This is consistent with the fitted initial delays for the single-qubit vacuum Rabi oscillations (main text \cref{fig:VacuumRabi}) in \cref{tab:simulationparams}. \par

Interestingly, the fitted initial delays for the state transfer (main text \cref{fig:StateTransfer}c) and accelerated vacuum Rabi oscillations (main text \cref{fig:Fock}) data are shorter, i.e., $t_0<$~\qty{1.5}{ns}.
We attribute this to the fact that in these experiments, 
the spin preparation was performed at a more negative detuning. For the single-qubit vacuum Rabi oscillations (main text \cref{fig:VacuumRabi}) and 2D state transfer (main text \cref{fig:StateTransfer}e), the spin is prepared at the readout point (see \cref{appendix:readout}).
Meanwhile, the other experiments (main text \cref{fig:StateTransfer}c, \cref{fig:Fock}) include an additional \qty{-3}{mV} detuning pulse to prepare the spin futher in the single-dot regime.
This means that during the ramp to the interaction point, considering that the ramp time is not changed, the system spends less time in the detuning range where the spin-photon interaction is enabled (see \cref{supfig:vrosim_vs_detuning}). \par

\clearpage
\section{Readout}
\label{appendix:readout}

\begin{figure*}[htbp] 
\center{\includegraphics[width=\linewidth]{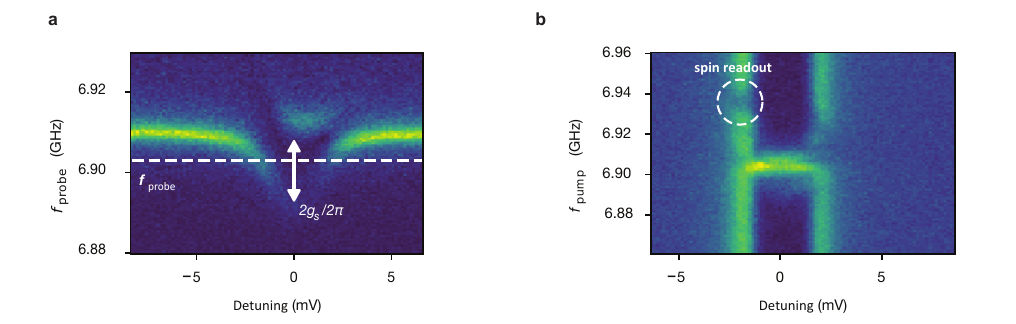}}
\caption{Coherent spin-photon coupling with detuning-pulsed dispersive readout.  \textbf{a.} Probed resonator frequency as function of double-dot detuning. At zero detuning, the charge degree of freedom of the double dot induces a dispersive shift of the resonator frequency. As the spin transition frequency is tuned into resonance with this charge-shifted cavity mode, spin-photon coupling leads to the observed vacuum Rabi splitting $2g_s/2\pi$. \textbf{b.} Two-tone spectroscopy at fixed probe frequency demonstrating quasi-dispersive spin sensing. As indicated in \textbf{a}, the probe frequency is fixed (at 6.903 GHz), such that any shift of the resonator frequency appears as a reduction in the measured signal-to-noise ratio. A pump tone at frequency $f_{\rm pump}$ is applied to a gate line to generate an excited spin-up population. The highlighted region shows a reduced SNR response, indicating that the presence of excited spin-up population induces a resonator frequency shift, consistent with a (quasi-)dispersive interaction.   }

\label{fig:Readout}
\end{figure*}

As discussed in the main text, achieving readout in the resonant regime is challenging due to the spin-photon hybridization. Therefore, we operate the device such that the spin qubit frequency can be tuned relative to the resonator mode by adjusting the interdot charge detuning. At zero charge detuning, the spin and resonator are brought into resonance while the charge degree of freedom remains off-resonant. In this situation, the eigenstates are hybridized combinations of spin and photon excitations~(main text \cref{fig:Readout}a), and coherent excitation exchange show as vacuum-Rabi oscillations as discussed in the main text. \par 

For spin readout, we adiabatically shift the charge detuning away from zero, but only slightly, such that the spin still retains partial spin–charge hybridization~(main text \cref{fig:Readout}b). The longitudinal magnetic-field gradient causes the spin transition frequency to shift with charge detuning, effectively moving the system into the dispersive regime of spin–photon coupling. In this regime, the spin and photon are no long resonant, but the spin state causes a state-dependent frequency shift on the resonator, which we detect through the transmitted microwave amplitude. This allows dispersive readout of the spin state while maintaining sufficient spin–photon coupling strength. Showing that the spin state can be read out after the resonant spin-photon interaction by pulsing the detuning.  \par

\end{document}